\documentclass[12pt]{article}
\pdfoutput=1
\usepackage{jheppub}

\usepackage{amsmath,amssymb,amsfonts}
\usepackage{graphicx}
\usepackage{color}
\usepackage{tikz}
\usepackage{soul}
\usepackage{dsfont}

\usepackage{booktabs}
\usepackage{multirow}

\usepackage{hyperref}

\usepackage{subfigure}

\usepackage{comment}

\makeatletter\def\l@subsubsection#1#2{}%
\makeatother

\def\be{\begin{eqnarray}}
\def\ee{\end{eqnarray}}

\def\makeatletter{\catcode`\@=11}
\makeatletter
\def\mathbox#1{\hbox{$\m@th#1$}}%
\def\math@ccstyles#1#2#3#4#5#6#7{{\leavevmode
      \setbox0\mathbox{#6#7}%
      \setbox2\mathbox{#4#5}%
      \dimen@ #3%
      \baselineskip\z@\lineskiplimit#1\lineskip\z@
      \vbox{\ialign{##\crcr
             \hfil \kern #2\box2 \hfil\crcr
             \noalign{\kern\dimen@}%
             \hfil\box0\hfil\crcr}}}}
\def\mathaccstyles{\math@ccstyles\maxdimen}
\def\maththroughstyles{\math@ccstyles{-\maxdimen}}
\def\unity%
 {\maththroughstyles{.45\ht0}\z@\displaystyle {\mathchar"006C}\displaystyle 1}

\title{Non-BPS branes and continuous symmetries}  

\author[a]{Oren Bergman,}
\author[b,c]{Eduardo Garcia-Valdecasas,}
\author[a]{Francesco Mignosa,}
\author[d]{Diego Rodriguez-Gomez}
\affiliation[a]{Department of Physics, Technion, Israel Institute of Technology, Haifa, 32000, Israel\\[-2mm]}
\affiliation[b]{SISSA, Via Bonomea 265, Trieste 34136, Italy\\[-2mm]}
\affiliation[c]{INFN, Sezione di Trieste, Via Valerio 2, 34127, Italy\\[-2mm]}
\affiliation[d]{Department of Physics, Universidad de Oviedo, C/Federico Garcia Lorca 18, 33007 Oviedo, Spain}
\affiliation[d]{Instituto Universitario de Ciencias y Tecnologias Espaciales de Asturias (ICTEA),C/ de la Independencia 13, 33004 Oviedo, Spain}

\abstract{We propose a holographic description of the operators implementing $U(1)$ global symmetries that are dual to superstring gauge fields in terms of non-BPS D-branes. We check the consistency of our proposal in a number of examples.}

\begin{document}

\maketitle

\flushbottom

\section{Introduction}

One of the key aspects of AdS/CFT duality, or {\em holography}, is that the global symmetries of the boundary theory
are dual to gauge symmetries in the bulk theory.
Recently we have seen that this ties in well with the more general understanding of symmetries in quantum field theory.
This is especially true in the case of discrete symmetries.
Discrete global symmetries are dual to discrete gauge fields in the bulk, which are in turn described in terms of
$U(1)$ gauge fields that couple via $BF$ terms \cite{Banks:2010zn}.  
This is closely related to the concept of the symmetry topological field  theory (SymTFT), a theory in one extra dimension that encodes the symmetries
of the QFT and their properties \cite{Freed:2012bs}.

\begin{figure}[h!]\label{fig.1}
	\centering    
	\subfigure[]{\label{Fig:NeumannBoundaryCondtions}\includegraphics[width=50mm]{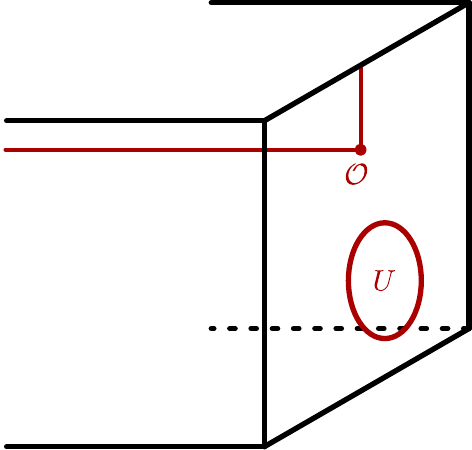}} \hspace{20mm}
	\subfigure[]{\label{Fig:DirichletBoundaryCondtions}\includegraphics[width=50mm]{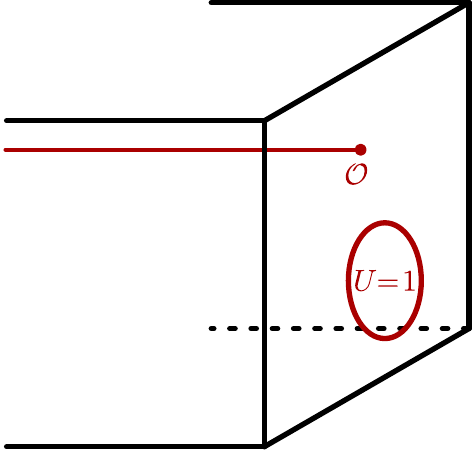}}
	\caption{(a) Neumann boundary conditions. Ending the bulk operator on the boundary gives a non-genuine charged operator $\mathcal{O}$. Placing the bulk operator on the boundary produces a non-trivial topological symmetry operator $U$. (b) Dirichlet boundary conditions. Ending the bulk operator on the boundary gives a genuine charged operator $\mathcal{O}$. Placing the bulk operator on the boundary produces a trivial operator  $U=1$.}
\end{figure}

The SymTFT of a QFT with discrete symmetries is generally a $BF$ theory, with possibly higher order topological terms
corresponding to anomalies.
Indeed the idea of a  SymTFT is strongly motivated by 
holography \cite{Witten:1998wy}.
For a QFT that admits a bulk dual the SymTFT is typically given by the near boundary IR limit of the bulk theory.
As shown in \cite{Witten:1998wy}, the SymTFT is actually associated to a class of QFT's (the so called {\em relative} QFT), and 
that to focus on a particular QFT (the {\em absolute} QFT) one has to specify boundary conditions on the gauge fields of the SymTFT. The symmetries of the absolute QFT and their properties are encoded in the spectrum of topological operators of the SymTFT. For a $BF$ type SymTFT the topological operators are given by the exponentials of the holonomies of the different gauge fields.
These may have to be modified due to anomalies.
Depending on the boundary conditions, a given topological operator of the SymTFT will
correspond in the boundary QFT either to a symmetry operator or to an operator on which a symmetry acts (a charged operator).
The former correspond to topological operators on submanifolds of the boundary, and the latter to topological operators on
manifolds that extend in the extra coordinate, and therefore must end on the boundary.
For example, for a one-form gauge field $A$ satisfying a Neumann boundary condition, the operator $\mathcal{O}= e^{i\int_C A}$, with $C$ in the boundary,
is a symmetry line-operator.
The would-be charged operator given by $e^{i\int_{C'} A}$, with $C'$ along the extra (radial) coordinate,
is not a genuine charged operator in the boundary theory, since it has the above line-operator attached to it.
On the other hand if $A$ satisfies a Dirichlet boundary condition, the operator at the boundary is trivial, and the operator ending on the boundary is genuine, see fig. \ref{fig.1}. 

The identification of the SymTFT as the bulk theory in holography also leads to the identification of the extended operators as the worldvolumes of branes.
It was understood from the early days of AdS/CFT that some charged operators are dual to branes or strings in the bulk,
whose worldvolumes extend in the radial direction.
This is the case when the global symmetry of the boundary theory is dual to a gauge field in the bulk that originates
from one of the gauge fields of ten-dimensional superstring theory (or eleven dimensional M theory). There are examples of charged operators for both discrete symmetries, like Wilson lines in ${\cal N}=4$ SYM that are dual to strings
ending on the boundary of $AdS_5$
\cite{Maldacena:1998im}, as well as for operators charged under a $U(1)$ symmetry, like baryons in Klebanov-Witten theory \cite{Klebanov:1998hh}, 
that are dual to D3-branes wrapping a 3-cycle in
the internal space and ending on the boundary \cite{Gubser:1998fp}.
More recently it has been understood that symmetry operators, at least for discrete symmetries, are also represented by branes, 
whose worldvolume is along the boundary \cite{OBtalks,Apruzzi:2022rei,GarciaEtxebarria:2022vzq,Bergman:2022otk,Heckman:2022muc}. The action of a symmetry operator on an operator that is charged under the corresponding symmetry is given by the link-pairing
between the two branes, which is read-off from the $BF$ term in the SymTFT.

For continuous symmetries the situation is different.
The analog of the SymTFT for a continuous symmetry is less understood, although there are some recent proposals in
\cite{Antinucci:2024zjp,Brennan:2024fgj,Bonetti:2024cjk,Apruzzi:2024htg}. Although operators that are charged under continuous symmetries are represented by branes in some cases,
as we mentioned above, there is no clear brane realization of the symmetry operators themselves.\footnote{See however \cite{Cvetic:2023plv}
for a recent proposal that continuous symmetry operators are dual to {\em fluxbranes} in the bulk.
A flux-$p$-brane is a localized supergravity 
solution with $(p+1)$-dimensional Poincare invariance and a nonzero $(D-p-1)$-form gauge field strength tangent to the transverse 
directions \cite{Gutperle:2001mb}.
However the worldvolume theory of a fluxbrane is not known.}
It is conceivable that continuous symmetries are simply not represented by branes.
A continuous symmetry operator can be expressed in terms of its conserved current.
In the case of interest the conserved current is explicitly given by the Hodge dual of the field strength of the bulk gauge field
under which the brane corresponding to the charged operator is charged.
Such currents are conserved by virtue of the Bianchi identity. 
The action of the symmetry operator on the charged operator then simply follows from the Gauss law in the bulk.

Nevertheless, we will offer an alternative proposal, which is that the symmetry operators for the type of continuous symmetries mentioned above 
are described holographically by {\em non-BPS branes}.
For each continuous symmetry of this type there exists a non-BPS, and unstable, brane in the bulk that links 
the charged BPS brane.
Crucially, we will argue that there is a continuous parameter associated to these branes and, therefore, that the symmetry implemented by them is continuous.
The dynamics of the tachyon on the non-BPS brane will clearly play an important role.

The rest of the paper is organized as follows.
In section 2 we review the tachyon effective theory of the non-BPS D-branes in Type II string theory,
and use it to argue that the non-BPS D-branes generate the (broken) global higher-form $U(1)$ symmetries of Type II string theory.
In section 3 we present our general proposal for $U(1)$ symmetries in QFT's that have a bulk holographic dual,
and provide three explicit examples: Klebanov-Witten theory, ABJM theory, 
and ${\cal N}=4$ SYM theory with $SU(N)$ gauge group. We then conclude and give some outlook for future directions.
 
\section{Non-BPS D-branes as broken symmetries}

Type II superstring theory has $U(1)$ $p$-form gauge fields.
Type IIA has odd-form gauge fields in the RR sector, and Type IIB has even-form gauge fields in the RR sector.
Both Type IIA and Type IIB have a 2-form gauge field, and its magnetic dual 6-form gauge field, in the NSNS sector.
In the absence of dynamical charged objects this would imply $U(1)$ global $(p+1)$-form symmetries generated
by the corresponding field strengths.
However these theories do have dynamical charged objects.
The fundamental string and the NS5-brane are charged under the NSNS gauge fields,
and the D-branes are charged under the RR gauge fields \cite{Polchinski:1995mt}.
Type IIA has charged D-branes of all even dimensions, and Type IIB has charged D-branes of all odd dimensions.
The charged D-branes are supersymmetric and therefore saturate the BPS bound relating their charge to their tension.
Polchinski also showed that the D-branes carry the minimal charge allowed by the Dirac quantization condition.
This implies that the global symmetries associated to the RR (and also NSNS) gauge fields are completely broken in string theory.
The currents are not conserved since the field strengths are not closed. There are source terms given by ``D-brane fields".
In the limit $g_s\rightarrow 0$ the D-branes have infinite tension, they become non-dynamical, and the symmetries are unbroken.

Type II string theory also has uncharged non-BPS D-branes.
In Type IIA they are odd dimensional and in Type IIB they are even dimensional.
The non-BPS D-branes are unstable to decay.
In the open string description, this instability manifests itself as a tachyonic mode.
The decay of a non-BPS D-brane corresponds to the condensation of the tachyon.\footnote{A similar process occurs in
the brane-anti-brane system for the charged BPS D-branes.} This process can be described in terms of the {\em tachyon effective action},
whose bosonic part is given by (we set $\alpha' =1$) \cite{Sen:2004nf}
\be
\label{TachyonAction}
 S_{\widetilde{\text{D}}p} =  \mbox{} &-& \int_{\Sigma_{p+1}} d^{p+1}x \, e^{-\Phi} V(T)\,\sqrt{\mbox{} -{\rm det}(G_{\mu\nu}+\partial_{\mu}T\partial_{\nu}T)} \nonumber\\
 &+&  \int_{\Sigma_{p+1}}  W(T)\,dT\wedge C_p \,,
\ee
where $\Phi, G_{\mu\nu}$ and $C_p$ are the background dilaton, metric, and RR $p$-form potential.\footnote{This action can be generalized in two ways. First, one can include the massless vector and scalar fields, as well as the background NSNS 2-form $B_2$. In particular in the WZ term $C_p$ is replaced by $C\wedge e^{\frac{1}{2\pi} (f_2 + B_2)}$. Second, one can include the dependence on the $\hat{A}$-genus. 
The tachyon effective action \eqref{TachyonAction} will generally suffice for us.}
The precise form of the functions $V(T)$ and $W(T)$ is not known, and in fact depends on what we mean precisely by the tachyon field $T$.
But we do know some of their properties:
\begin{enumerate}
\item Both $V(T)$ and $W(T)$ are positive even functions of $T$ that approach zero asymptotically as $e^{-|T|/\sqrt{2}}$.
\item $V(0) = \tilde{T}_p$ is the tension of the non-BPS D$p$-brane, which is related to the tension of the BPS D$p$-brane
by $\tilde{T}_p = \sqrt{2} \, {T}_p$.
\item The identification of the tachyon kink in the non-BPS D$p$-brane as the BPS D$(p-1)$-brane requires
\be
\int_{-\infty}^{\infty} dT \, V(T) = \int_{-\infty}^{\infty} dT \, W(T) = {T}_{p-1} \,.
\ee
\end{enumerate}
In no sense is this a {\em low-energy} effective theory, since the tachyon is not light compared to the string scale.
It is an effective theory in the sense that solutions of its equation of motion reproduce certain worldsheet BCFT's involving the tachyon.\footnote{
An explicit function with these properties, and also reproducing the rolling tachyon BCFT as a classical solution, is given by 
$V(T) = W(T) = \tilde{T}_p/\cosh(T/\sqrt{2})$.}

The action appears to vanish in the tachyon vacuum at $T\rightarrow \pm \infty$.
This is consistent with the decay of the non-BPS D-brane, which eliminates the open string degrees of freedom \cite{Sen:1999md}.
But we should be more careful about this conclusion because of the coupling to the background RR potential $C_p$,
which is not in general well defined.
The field strength $F_{p+1} = dC_p$ is well defined, so we can try to remedy the situation by integrating the WZ term by parts.
Dropping the boundary term, this gives \cite{Cornalba:2005jd},
\be
S_{\text{WZ}} = \mbox{} -  \int_{\Sigma_{p+1}} Z(T) \, F_{p+1}
\ee
where 
\be
Z(T) = \int_{T_0}^T W(T^\prime) dT^\prime \,.
\ee
Note that there is an arbitrary constant of integration depending on the choice of $T_0$. The action in the tachyon vacuum is therefore
\be
\label{TachyonVacuumAction}
S^{\text{vac}}_{\widetilde{\text{D}}p} = S_{\widetilde{\text{D}}p}[T\rightarrow \pm \infty] = {\alpha\over 2\pi}  \int_{\Sigma_{p+1}} F_{p+1}
\ee
where $\alpha$ is an arbitrary dimensionless constant. 
The quantization of RR flux implies a periodicity $\alpha \sim \alpha + 2\pi$.\footnote{We normalize the RR fields to have
flux quantized in integer multiples of $2\pi$.} This appears to be a new {\em physical} parameter in the worldvolume theory of a non-BPS 
D$p$-brane. 

One should contrast this with the BPS D$p$-brane.
There is an analogous ambiguity when integrating by parts the 
WZ term $C_{p-1}\wedge f_2$ to get $F_p \wedge a$.
But in this case it corresponds to a large gauge transformation of the worldvolume gauge field $a$.
The quantization of the D$p$-brane charge ensures that the action is invariant, so this has no physical effect.

So we find that there is a remnant of the non-BPS D$p$-brane even after it has decayed, and this remnant is proportional to the
RR $(p+1)$-form flux in the worldvolume of the $p$-brane.
This is the flux sourced by a BPS D$(7-p)$-brane.
Therefore a non-BPS D$p$-brane surrounding a BPS D$(7-p)$-brane measures its charge via Gauss' law,
and thereby implements the global $U(1)$ transformation acting on the $(8-p)$-dimensional operator
corresponding to the worldvolume of the D$(7-p)$-brane. But as we said, these global symmetries are explicitly broken (except at $g_s=0$) since D-branes are dynamical objects in string theory. The operator given by exponentiating the action in \eqref{TachyonVacuumAction} is not topological.


\section{Non-BPS branes as unbroken symmetries, with examples}

In a holographic background the symmetry may be unbroken at the boundary.
This is what happens for example to discrete symmetries generated by BPS D-branes that are taken to the boundary of the bulk spacetime.
In this limit the D-brane action becomes topological, since the DBI term vanishes.
For the non-BPS D-brane the action \eqref{TachyonVacuumAction} becomes topological at the boundary
since the BPS D-brane sourcing the RR field has an infinite tension at its endpoint on the boundary.

\begin{figure}[h!]
 \centering  \includegraphics[width=0.7\textwidth]{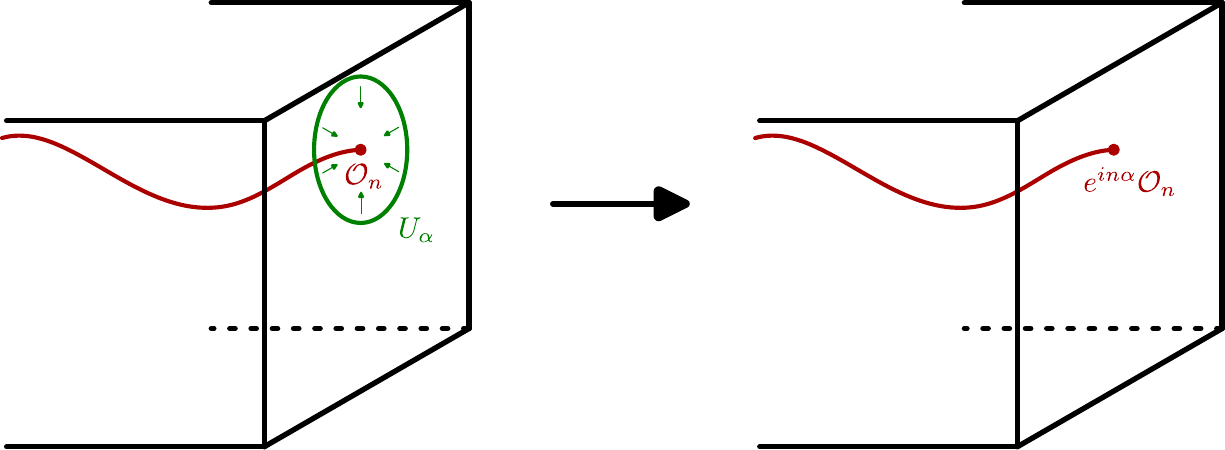}   \caption{A non-BPS D-brane located at the boundary becomes a topological symmetry operator $U_\alpha$ that acts on a D-brane ending on the boundary $\mathcal{O}_n$ by the linking action \eqref{Eq:LinkingAction}. }  	 \label{Fig:BoundaryAction}
\end{figure}

Our general proposal is as follows.
Consider a $d$-dimensional SCFT that is dual to Type II string theory on $AdS_{d+1} \times M_{9-d}$,
with a $U(1)$ $p$-form gauge field originating from the reduction on a $q$-cycle $\Sigma_q \subset M_{9-d}$ of 
a ten-dimensional RR $(p+q)$-form gauge field.
Imposing Dirichlet boundary conditions on this gauge field implies that the SCFT has a $U(1)$ global $(p-1)$-form symmetry.
The operators charged under this symmetry are dual to D$(p+q-1)$-branes wrapping $\Sigma_q$ and extending in the radial direction.
The charge $n$ operator is explicitly given by
\be
\label{ChargedOperator}
{\cal O}_n[M_{p-1}] = e^{inS_{\text{D}(p+q-1)}} = 
e^{in\int_{\Sigma_q \times M_{p-1}\times \mathbb{R}_+} C_{p+q}} \,.
\ee
Our new proposal is that the symmetry operators correspond to a non-BPS $\widetilde{\text{D}}(8-p-q)$-brane wrapping the dual $(9-d-q)$-cycle
$\Sigma_{9-d-q} \subset M_{9-d}$, and transverse to the radial direction.
These are given explicitly by
\be
\label{SymmetryOperator}
U_\alpha[M_{d-p}]  =  e^{i S^{\text{vac}}_{\widetilde{\text{D}}(8-p-q)}} = 
e^{i{\alpha\over 2\pi} \int_{\Sigma_{9-d-q}\times M_{d-p}} F_{9-p-q}} \,.
\ee
If we take the space $M_{d-p}$, which the non-BPS brane occupies, to be a closed $(d-p)$-dimensional submanifold of $AdS_{d+1}$ that surrounds the 
$(p-1)$-dimensional submanifold $M_{p-1}$, which the BPS branes occupy, then Gauss' law tells us that 
\be
 \oint_{\Sigma_{9-d-q}\times M_{d-p}} F_{9-p-q} = 2\pi n \,.
\ee
The action of the operator in \eqref{SymmetryOperator} on the operator in \eqref{ChargedOperator} is therefore given by
\be 
U_\alpha[M_{d-p}] \, {\cal O}_n[M_{p-1}] = e^{in\alpha} \, {\cal O}_n[M_{p-1}] \,, \label{Eq:LinkingAction}
\ee
which is precisely the action of the $U(1)$ $(p-1)$-form symmetry, see fig. \ref{Fig:BoundaryAction}.
We now give a few examples of this in some well known holographic theories.

\subsection{Klebanov-Witten theory}
\label{baryon}

Our first example is Klebanov-Witten theory, a 4d ${\cal N}=1$ supersymmetric gauge theory with gauge group $SU(N)\times SU(N)$,
a pair of chiral multiplets $A_i$ in the $({\bf N},\bar{\bf N})$ representation, a pair $B_i$ in the $(\bar{\bf N},{\bf N})$ representation,
and a certain superpotential coupling between them \cite{Klebanov:1998hh}.
This theory is obtained as the worldvolume theory of $N$ D3-branes probing a conifold singularity in Type IIB string theory.
The large $N$ dual is Type IIB string theory on $AdS_5 \times T^{1,1}$, where $T^{1,1}$ is the 5d base of the conifold,
with $N$ units of RR 5-form flux.

This 4d gauge theory has a baryonic $U(1)$ global symmetry under which the fields $A_i$ carry charge $+1$ and the fields
$B_i$ carry charge $-1$. There are gauge-invariant baryon-like operators ${\cal B}$ given by $N$-fold products of the chiral superfields
that are antisymmetrized with respect to the gauge groups \cite{Gubser:1998fp}. Upon appropriate normalization, they carry $\pm 1$ unit of charge. 
The baryonic $U(1)$ symmetry is dual to a $U(1)$ gauge field in $AdS_5$ that comes from reducing the RR
4-form potential $C_4$ on the 3-cycle $S^3\subset T^{1,1}$, and the baryon operators are dual to a D3-brane wrapping $S^3$,
extending in the radial direction, and ending on the boundary of $AdS_5$.

Our proposal for the brane corresponding to the baryonic $U(1)$ symmetry operator is a non-BPS D4-brane that wraps the 2-cycle $S^2\subset T^{1,1}$,
extends along a 3-manifold $M_3$ in spacetime, and is taken to the boundary of $AdS_5$.
The $U(1)$ symmetry operator is therefore given by 
\be
U_\alpha[M_3] = e^{i S^{\text{vac}}_{\widetilde{\text{D}}4}} = e^{i{\alpha\over 2\pi} \int_{S^2 \times M_3} F_5} \,. 
\ee
The $n$-baryon operator is given by
\be
{\cal B}_n = e^{in\int_{S^3 \times \mathbb{R}_+} C_4} \,.
\ee
Taking $M_3$ to be a closed 3-manifold surrounding ${\cal B}_n$ gives the $U(1)$ action:
\be
U_\alpha[M_3] \, {\cal B}_n =  e^{in\alpha} \, {\cal B}_n \,.
\ee

\subsection{ABJM theory}

ABJM theory is a 3d ${\cal N}=6$ supersymmetric Chern-Simons-Matter theory with gauge group $U(N)_k\times U(N)_{-k}$
and matter in the bi-fundamental representation \cite{Aharony:2008ug}.
This theory has a non-trivial $U(1)$ global symmetry generated by the diagonal monopole number current $J = *\mbox{Tr} (F_1 + F_2)$.
This acts on diagonal monopole operators, that are dressed with matter fields into gauge invariant operators.
The holographic dual of ABJM theory can be described either as M theory on $AdS_4\times S^7/\mathbb{Z}_k$ with $N$ units of 7-form flux, 
or Type IIA string theory
on $AdS_4\times \mathbb{C}P^3$ with $N$ units of RR 6-form flux and $k$ units of RR 2-form flux.
We will use the Type IIA description (which strictly speaking is valid for $N^{1/5} \ll k\ll N$).

The $U(1)$ global symmetry is dual to a $U(1)$ gauge field in $AdS_4$ given by the RR 1-form $C_1$,
which is taken to satisfy a Dirichlet boundary condition.\footnote{There are in fact two gauge fields in the bulk, with one linear combination 
being Higgsed \cite{Aharony:2008ug}. In the present case, the second gauge field satisfies a Neumann boundary condition.
Other boundary conditions are possible, and lead to variants of ABJM theory \cite{Bergman:2020ifi} which we will not discuss here.}
The objects charged under $C_1$ are D0-branes, therefore
the dressed monopole operator corresponds to a Euclidean D0-brane whose worldline ends on the boundary.

The $U(1)$ symmetry operator in this case is described by a non-BPS D7-brane wrapping the entire internal space $\mathbb{C}P^3$,
extending in two spacetime directions, and taken to the boundary of $AdS_4$. The symmetry operator is therefore given by 
\be
U_\alpha[M_2] = e^{i S^{\text{vac}}_{\widetilde{\text{D}}7}} = e^{i{\alpha\over 2\pi} \int_{\mathbb{C}P^3 \times M_2} F_8} \,. 
\ee
 The $n$-monopole operator is given by
\be
{\cal M}_n = e^{in\int_{\mathbb{R}_+} C_1} \,.
\ee
Taking $M_2$ to be a closed 2-manifold surrounding ${\cal M}_n$ gives
\be
U_\alpha[M_2] \, {\cal M}_n =  e^{in\alpha} \, {\cal M}_n \,.
\ee

There are other non-BPS D-branes that one can consider, but they act unfaithfully in the ABJM theory.
For example a non-BPS D3-brane wrapping the 2-cycle $\mathbb{C}P^1\subset \mathbb{C}P^3$ couples to $F_4$ and therefore
links the BPS D4-brane wrapping the 4-cycle $\mathbb{C}P^2$.
However, given the boundary conditions dual to the $U(N)\times U(N)$ theory,
the wrapped D4-brane corresponds to a gauge-non-invariant operator, the di-baryon \cite{Bergman:2020ifi}.
In other variants of ABJM theory it will be a gauge-invariant operator, and the non-BPS D3-brane will act faithfully.

\subsection{${\cal N}=4$ SYM theory with $SU(N)$ gauge symmetry}

${\cal N}=4$ Super-Yang-Mills theory with gauge group $SU(N)$ is dual to Type IIB string theory on $AdS_5\times S^5$ \cite{Maldacena:1997re}.
This theory does not really have a continuous global symmetry. 
There is only a discrete $\mathbb{Z}_N$ 1-form symmetry acting on Wilson lines.
In the bulk this corresponds to the link-pairing of D1-brane worldsheets at the boundary and fundamental string worldsheets ending on the boundary.

However, in a sense, the Yang-Mills $\theta$-parameter may be regarded as a $U(1)$ ``$(-1)$-form symmetry".
$(-1)$-form symmetries are very different from ordinary 0-form and higher $p$-form symmetries.
A $(-1)$-form symmetry is not associated with a conserved current or a topological operator in the usual sense \cite{Aloni:2024jpb}. 
No operators are charged under a $(-1)$-form symmetry, and it does not give rise to selection rules. 
The physical meaning of a $(-1)$-form symmetry is limited to the statement that the path integral is a sum over sectors classified 
by the $(-1)$-form symmetry charge. 
One can, however, couple a $(-1)$-form symmetry to a background field, gauge it, and study its anomalies \cite{Cordova:2019jnf,Cordova:2019uob,Choi:2022odr,Brennan:2024tlw}.

From the point of view of holography, however, $(-1)$-form symmetries are quite similar to higher form symmetries.
A $U(1)$ $(-1)$-form symmetry in a holographic QFT is dual to a 0-form gauge field, namely to a compact scalar field, in the bulk.
In the present case this is the Type IIB RR scalar $C_0$, which is $2\pi$ periodic, $C_0 \sim C_0 + 2\pi$.
The boundary value of $C_0$ is identified with the Yang-Mills $\theta$ parameter.
The action of the $(-1)$-form symmetry is to shift $\theta$ by a constant\footnote{Note that $\theta$ can be understood as a background gauge field for the $(-1)$-form symmetry, so eq. \eqref{Eq:-1action} is strictly speaking the action on the background gauge field.}
\be 
\label{ThetaShift}
\theta \rightarrow \theta + \alpha \,, \label{Eq:-1action}
\ee
where $\alpha \in [0,2\pi)$.
This is not really a symmetry since it changes the theory.
In the bulk this is implemented by a non-BPS D8-brane which wraps $S^5$ and extends in all four spacetime coordinates.
The symmetry operator is given by
\be
\label{D8Operator}
U_\alpha[M_4] = e^{iS^{\text{vac}}_{\widetilde{\text{D}}8}} = e^{i{\alpha\over 2\pi}\int_{S^5\times M_4} F_9} \,.
\ee
But there is no corresponding charged operator in the boundary theory. The D-brane sourcing $F_9$ is a D$(-1)$-brane. Since the D$(-1)$-brane has no worldvolume, obviously it cannot end on the boundary.
How then can we understand the action in \eqref{ThetaShift}?
The D$(-1)$-brane corresponds to a local operator in the bulk, which the D8-brane can link.
For $n$ D$(-1)$-branes this operator is $e^{in C_0}$, and then by Gauss' law the action of the D8-brane is given by
\be
e^{i{\alpha\over 2\pi}\int_{S^5\times M_4} F_9} \, e^{in C_0} = e^{i n \alpha} \, e^{in C_0} \,.
\ee
In other words the RR scalar is shifted by $\alpha$
\be
\label{C0Shift}
C_0 \rightarrow C_0 + \alpha \,,
\ee
which corresponds to \eqref{ThetaShift} in the boundary theory.

Equation (\ref{C0Shift}) is just the 0-form symmetry of Type IIB string theory associated to the 0-form gauge field $C_0$.
As with all the higher form symmetries, the 0-form symmetry is also broken.
In this case it is because the D$(-1)$-branes are dynamical.
This manifests itself in the fact that the string theory path integral includes a sum over D$(-1)$-branes,
where the $n$ D$(-1)$-brane term is weighted by the phase $e^{in C_0}$ \cite{Green:1997tv}. 
In fact this persists also at the boundary, since the D$(-1)$-brane action remains finite.
This is of course consistent with the fact that the action on $\theta$ (\ref{ThetaShift}) is not a symmetry of the field theory.

\section{Conclusions and future directions}

In this note we have proposed a novel string theory realization of $U(1)$ symmetry operators in holographic theories in terms of bulk non-BPS D-branes. The key observation is that the worldvolume theory of non-BPS branes admits a continuous periodic parameter $\alpha$. In the tachyon vacuum, the worldvolume action of the non-BPS D-brane has a remnant proportional to $\alpha$ and to an RR field strength,
such that when taken to the holographic boundary, the operator defined by exponentiating the action becomes topological. 
This operator links non-trivially with the BPS D-brane that sources the RR field strength, and therefore acts on the operator
dual to the BPS D-brane as a $U(1)$ transformation with parameter $\alpha$.

We have demonstrated our proposal in a number of examples. These can be generalized in many ways.
In particular, baryonic $U(1)$ symmetries and monopole-like $U(1)$ symmetries associated to Bianchi identities are generic ingredients of 
holographic theories. One should be able to match every such symmetry with a particular non-BPS brane in the dual background.
Analogous symmetries are also abundant in higher-dimensional holographic theories.
Furthermore, for a given bulk theory there are different possible boundary theories depending on the boundary conditions for the bulk gauge fields.
Among the several non-BPS branes in the bulk, some will correspond to faithful symmetries that act non-trivially on genuine charged operators,
and others will not. It would be interesting to study this in detail, for example in the ABJM-like theories \cite{Bergman:2020ifi}.

It is well-known that non-BPS D$p$-branes can also emerge as tachyon kinks in a $\text{D}(p+1)-\overline{\text{D}(p+1)}$ brane system. It would be interesting to explore whether this description can give us additional insights into continuous symmetry operators, such as their abundance in holographic backgrounds and their universal properties. Also, coincident non-BPS D-branes admit Myers-like  couplings \cite{Myers:1999ps,Janssen:2000sz}, hinting potential non-trivial fusion rules signaling non-invertible symmetries. Both the tachyon condensation on the $\text{D}(p+1)-\overline{\text{D}(p+1)}$ system and the Myers effect might lead to non-trivial relations among different non-BPS branes in a given holographic set-up which would be very interesting to explore.

\section*{Acknowledgements}

We thank Matteo Bertolini and Marco Serone for insightful conversations. The work of O.B. and F.M. is supported in part by the Israel Science Foundation under grant No. 1254/22, and by the US-Israel Binational Science Foundation under grant No. 2022100. The work of EGV has been partially supported by Margarita Salas award CA1/RSUE/2021-00738, MIUR PRIN Grant 2020KR4KN2 and by INFN Iniziativa Specifica ST\&FI. D.R.G is supported in part by the Spanish national grant MCIU-22-PID2021-123021NB-I00.

\bibliography{nonBPSbib}
\bibliographystyle{JHEP}

\end{document}